\documentclass[aps,preprint,nofootinbib]{revtex4}%
\usepackage{amsfonts}
\usepackage{amsmath}
\usepackage{amssymb}
\usepackage{graphicx}%
\providecommand{\U}[1]{\protect\rule{.1in}{.1in}}

\begin{document}
\preprint{ }
\title[Short title for running header]{Comment on `Hamiltonian formulation for the theory of gravity and canonical
transformations in extended phase space' by T P Shestakova}
\author{N Kiriushcheva}
\email{nkiriush@uwo.ca}
\author{P G Komorowski}
\email{pkomoro@uwo.ca}
\author{S V Kuzmin}
\email{skuzmin@uwo.ca}
\affiliation{The Department of Applied Mathematics, The University of Western Ontario,
London, Ontario, N6A 5B7, Canada}

\begin{abstract}
We argue that the conclusion, `we cannot consider the Dirac approach as
fundamental and undoubted', made in the paper by Shestakova (\textit{Class.
Quantum Grav. }\textbf{28 }055009, 2011), is based upon an incomplete and
flawed analysis of the simple model presented in section 3 of the article. We
re-examine the analysis of this model and find that it does not support the
author's conclusion. For the theory of gravity neither the equivalence of the
effective action nor its Hamiltonian formulation is given by the author,
therefore, we only provide a brief commentary.

\end{abstract}
\volumeyear{year}
\volumenumber{number}
\issuenumber{number}
\eid{identifier}
\date{\today}
\received{}

\maketitle


We examine the analysis of the simple Lagrangian, which was used in section 3
of \cite{ShestakovaCQG} to illustrate that `[Dirac's] algorithm fails to
produce correct results'\ for an arbitrary parametrisation. Two
parametrisations (equations (S12) and (S13))\footnote{Equations indicated as
(S\#\#) are from \cite{ShestakovaCQG}.} are discussed:%

\begin{equation}
L_{1}=-\frac{1}{2}\frac{a\dot{a}^{2}}{N}+\frac{1}{2}Na\text{ };\text{
\ \ \ \ \ }L_{2}=-\frac{1}{2}\frac{a\dot{a}^{2}}{\sqrt{\mu}}+\frac{1}{2}%
\sqrt{\mu}a,\text{ \ \ \ \ \ }N=\sqrt{\mu}. \label{eqnS10}%
\end{equation}

In \cite{ShestakovaCQG}, Dirac's algorithm\footnote{When we refer to Dirac's
algorithm, we mean that all steps are performed: from introducing momenta for
\textit{all }variables, to finding gauge invariance using, for example,
Castellani's algorithm \cite{Castellani}, based on the Dirac conjecture that
all of the first class constraints are needed to derive gauge transformations
\cite{Diracbook}.} is applied to $L_{1}$ and $L_{2}$ to build the gauge
generators and then to find the corresponding gauge symmetries. For $L_{2}$,
the generator, (S26), gives%

\begin{equation}
\delta_{2}\mu=-\frac{1}{2\mu}\dot{\mu}\theta_{2}+\dot{\theta}_{2}\text{
},\text{\ \ \ \ \ }\delta_{2}a=\frac{1}{2}\frac{\dot{a}}{\mu}\theta
_{2}~;\label{eqnS14}%
\end{equation}
where $\theta_{2}$ is a gauge parameter. In addition to $\delta_{2}\mu$ (see
equation (S27)), which was deemed in \cite{ShestakovaCQG} to be the correct
transformation, we used equation (S26) to obtain $\delta_{2}a$. By applying
the same method to $L_{1}$, another generator is found, (S32), which leads to
the following transformations:%

\begin{equation}
\delta_{1}N=\dot{\theta}_{1}\text{ },\text{\ \ \ }\delta_{1}a=\frac{\dot{a}%
}{N}\theta_{1}\text{\ },\text{\ } \label{eqnS15}%
\end{equation}
which were not reported in \cite{ShestakovaCQG}. Instead, it was declared that
Dirac's method does not produce a `correct' result for the $L_{1}$
parametrisation, and must be abandoned in favour of another method, the
Extended Phase Space (EPS) approach.

Transformations (\ref{eqnS14}) and (\ref{eqnS15}) are written for different
variables; therefore, to compare them we shall use $N=\sqrt{\mu}$. The
transformations of the fields in $L_{2}$ under $\delta_{1}$ are%

\begin{equation}
\delta_{1}\mu=2N\delta_{1}N=2\sqrt{\mu}\dot{\theta}_{1}\text{ }%
,\text{\ \ \ \ }\delta_{1}a=\frac{\dot{a}}{\sqrt{\mu}}\theta_{1}%
~;\label{eqnS16}%
\end{equation}
similarly, the transformations of the fields of $L_{1}$ under $\delta_{2}$ are%

\begin{equation}
\delta_{2}N=\delta_{2}\sqrt{\mu}=\frac{1}{2\sqrt{\mu}}\delta_{2}\mu=-\frac
{1}{2N^{2}}\dot{N}\theta_{2}+\frac{1}{2N}\dot{\theta}_{2}=\left(  \frac{1}%
{2N}\theta_{2}\right)  _{,0}\text{ },\text{\ \ \ \ }\delta_{2}a=\frac{1}%
{2}\frac{\dot{a}}{N^{2}}\theta_{2}~.\text{\ \ } \label{eqnS17}%
\end{equation}
Hence transformations (\ref{eqnS14}) and (\ref{eqnS15}) are different for both
sets of fields (compare (\ref{eqnS14}) with (\ref{eqnS16}) and (\ref{eqnS15})
with (\ref{eqnS17})).

If transformation (\ref{eqnS14}) (the `correct' one) is applied to $N$, we
obtain (\ref{eqnS17}); this variation differs from equation (S33), which is
reported in \cite{ShestakovaCQG} to be the expected result. Let us designate
equation (S33) as the `second correct' result, and associate it with a
transformation $\delta_{3}$.

What is the meaning of these various transformations? For a Lagrangian with a
gauge symmetry, according to Noether's second theorem \cite{Noether}, there is
a corresponding combination of Euler-Lagrange derivatives (ELD) -- a
differential identity (DI). If a transformation is known, the corresponding DI
can always be restored \cite{Schwinger}. For example, a DI for $L_{2}$ can be
found from%

\begin{equation}
\int\left[  E_{\left(  \mu\right)  }^{\left(  2\right)  }\delta_{2}%
\mu+E_{\left(  a\right)  }^{\left(  2\right)  }\delta_{2}a\right]  dt=\int
I^{\left(  2\right)  }\theta_{2}dt~, \label{eqnS18}%
\end{equation}
where $E_{\left(  \mu\right)  }^{\left(  2\right)  }=\frac{\delta L_{2}%
}{\delta\mu}$ and $E_{\left(  a\right)  }^{\left(  2\right)  }=\frac{\delta
L_{2}}{\delta a}$ are ELDs of $L_{2}$. Substituting the known gauge
transformations (\ref{eqnS14}) and performing simple rearrangements, we obtain%

\begin{equation}
I^{\left(  2\right)  }=-\frac{1}{2\mu}\dot{\mu}E_{\left(  \mu\right)
}^{\left(  2\right)  }-\dot{E}_{\left(  \mu\right)  }^{\left(  2\right)
}+\frac{1}{2}\frac{\dot{a}}{\mu}E_{\left(  a\right)  }^{\left(  2\right)
}\equiv0. \label{eqnS20}%
\end{equation}
Similarly, for $L_{1}$ the DI is%

\begin{equation}
I^{\left(  1\right)  }=-\dot{E}_{\left(  N\right)  }^{\left(  1\right)
}+\frac{\dot{a}}{N}E_{\left(  a\right)  }^{\left(  1\right)  }\equiv0.
\label{eqnS22}%
\end{equation}

These results can be directly verified by substituting the corresponding ELDs
or by performing transformations of the Lagrangian (e.g. $\delta_{1}%
L_{1}=\partial_{0}\left(  -\frac{a\dot{a}^{2}}{2N^{2}}\theta_{1}+\frac{1}%
{2}a\theta_{1}\right)  $); thus confirming that Dirac's algorithm correctly
finds a symmetry of the Lagrangian. In the Lagrangian approach, if one DI is
found (e.g. using Dirac's algorithm), we can build more DIs and find new gauge
symmetries by repeating steps (\ref{eqnS18})-(\ref{eqnS20}) in inverse order.
For example, let us modify DI (\ref{eqnS20})%

\begin{equation}
\tilde{I}^{\left(  2\right)  }=2\sqrt{\mu}I^{\left(  2\right)  }=-\partial
_{0}\left(  2\sqrt{\mu}E_{\left(  \mu\right)  }^{\left(  2\right)  }\right)
+\frac{\dot{a}}{\sqrt{\mu}}E_{\left(  a\right)  }^{\left(  2\right)  }\equiv0;
\label{eqnS26}%
\end{equation}
the transformations that this DI produces are the same as (\ref{eqnS16}), so
this is also symmetry of $L_{2}$. Similarly, considering%

\begin{equation}
\tilde{I}^{\left(  1\right)  }=\frac{1}{2N}I^{\left(  1\right)  }=-\frac
{1}{2N}\dot{E}_{\left(  N\right)  }^{\left(  1\right)  }+\frac{\dot{a}}%
{2N^{2}}E_{\left(  a\right)  }^{\left(  1\right)  }\equiv0, \label{eqnS28}%
\end{equation}
we obtain transformations (\ref{eqnS17}). So, symmetries\ (\ref{eqnS14}) and
(\ref{eqnS17}) for the `correct' expressions and those for the `incorrect'
expressions, (\ref{eqnS15}) and (\ref{eqnS16}), are symmetries for both
Lagrangians. More symmetries can be found by further modification of the DIs;
and many parametrisations of a Lagrangian can be explored. For any symmetry
specified, we can find a parametrisation for which Dirac's algorithm will lead
to this same symmetry; e.g. for the `second correct' symmetry the
parametrisation is:%

\begin{equation}
N=e^{-\varkappa}\text{ },\text{\ \ \ \ }L_{3}=-\frac{1}{2}e^{\varkappa}%
a\dot{a}^{2}+\frac{1}{2}e^{-\varkappa}a~. \label{eqnS29}%
\end{equation}
Repeating Dirac's analysis, as was done in \cite{ShestakovaCQG} for
(\ref{eqnS10}), one obtains:%

\begin{equation}
\delta_{3}\varkappa=-\dot{\varkappa}\theta_{3}+\dot{\theta}_{3}~,\text{
\ \ \ \ }\delta_{3}a=-\dot{a}\theta_{3}\text{ \ \ \ and \ }\delta_{3}%
N=-\dot{N}\theta_{3}-N\dot{\theta}_{3}~,\text{ \ \ \ \ }\delta_{3}a=-\dot
{a}\theta_{3}~.\text{ } \label{eqnS30}%
\end{equation}
So, the parametrisation of $L_{3}$ (not $L_{2}$) leads to equation (S33) --
the `second correct' symmetry.

The justification to call transformations (\ref{eqnS14}), from the application
of Dirac's method to $L_{2}$, `correct' is based on an `interpretation' of the
field $\mu$ as the component $g_{00}$ of the metric tensor and on its
invariance under diffeomorphism (see (S28), (S29)), which is known from the
Einstein-Hilbert (EH) (not $L_{2}$) action; the components of the vector gauge
parameter $\theta^{\lambda}$ must be carefully crafted: $\theta^{0}%
=\frac{\theta_{2}}{2\mu}$ and $\theta^{k}=0$. \ This approximation must be
applied to all fields of a given model if one expects this `diffeomorphism' to
be a symmetry of $L_{2}$. The transformation of a scalar under diffeomorphism
is known; and the same approximation leads to%

\begin{equation}
\delta_{\mathit{diff}}a=-a\partial_{\mu}\theta^{\mu}\Longrightarrow\delta
a=-a\partial_{0}\left(  \frac{\theta_{2}}{2\mu}\right)  =\frac{a\dot{\mu}%
}{2\mu^{2}}\theta_{2}-\frac{a}{2\mu}\dot{\theta}_{2}~.\label{eqnS31}%
\end{equation}
For $a$, there are no time derivatives of the gauge parameters in
(\ref{eqnS14}), (\ref{eqnS15}), or in (\ref{eqnS30}); therefore, the rationale
for choosing this `correct' transformation is based on a questionable
`interpretation' and `approximation', which is not internally consistent.

The three examples considered here illustrate the equivalence of the
Lagrangian and Hamiltonian methods for systems with gauge invariance, and show
that all Lagrangian symmetries can also be derived using the Hamiltonian
approach. The failure to find a parametrisation (as $L_{3}$) to derive a
particular symmetry in the Hamiltonian approach is not a failure of Dirac's
method, and it is not a strong enough justification to advocate the use of a
new approach: EPS or any other. For these examples, all symmetries can be
derived in both approaches (Lagrangian and Hamiltonian). The question of which
symmetry is `correct', is beyond the realm of Lagrangian and Hamiltonian
equivalence; it cannot be answered by performing a `canonization' of a
symmetry one expects\footnote{There may be a specific exception.\ In covariant
theories one should expect a covariant result and expect that a covariant
parametrisation would be preferable for the Hamiltonian, or as in the example
considered, due to the simplicity of its Lagrangian.}; a mathematical
criterion is required. \ 

In the case of the EH action we analysed the group properties of various
symmetries \cite{KKK-1}. Let us do the same for this model by calculating the
commutators of different transformations. For $\delta_{1}$, the
transformations which were not included in \cite{ShestakovaCQG} (perhaps
because they were deemed incorrect), we obtain%

\begin{equation}
\left[  \delta_{1}^{\prime\prime},\delta_{1}^{\prime}\right]  =\left(
\delta_{1}^{\prime\prime}\delta_{1}^{\prime}-\delta_{1}^{\prime}\delta
_{1}^{\prime\prime}\right)  \left(  \left(  N,a\right)  ,\left(  \mu,a\right)
,\left(  \varkappa,a\right)  \right)  =0. \label{eqnS35}%
\end{equation}
This is the simplest possible result (as in the Maxwell theory). For
$\delta_{2}$, the `correct' transformations, the result is%

\begin{equation}
\left[  \delta_{2}^{\prime\prime},\delta_{2}^{\prime}\right]  \left(
\mu,a\right)  =\delta_{2}^{\prime\prime\prime}\left(  \mu,a\right)  ,\text{
\ \ \ \ }\theta_{2}^{^{\prime\prime\prime}}=\frac{1}{2\mu}\left[  \theta
_{2}^{\prime\prime}\dot{\theta}_{2}^{\prime}-\theta_{2}^{\prime}\dot{\theta
}_{2}^{\prime\prime}\right]  ,\label{eqnS38}%
\end{equation}
which has a field-dependent `soft algebra'\ structure\footnote{For the other
two pairs, the field dependence is different, but consistent with field
redefinitions $\left(  N,a\right)  $ and $\left(  \varkappa,a\right)  $
:\ \ $\frac{1}{2N^{2}}$ \ and $\ \frac{1}{2}e^{2\varkappa}$ in (\ref{eqnS38}),
instead of $\frac{1}{2\mu}$.}. In such a case it might be possible that the
Jacobi identity is not satisfied (i.e. failure to form a group); to check
this, the evaluation of double commutators is needed, as was performed for the
ADM transformations in \cite{KKK-1}. Group properties notwithstanding, this
`correct' symmetry (\ref{eqnS38}) is more complicated than the `incorrect' one
(\ref{eqnS35}). For $\delta_{3}$, the `second correct' symmetry, the
commutator is%

\begin{equation}
\left[  \delta_{3}^{\prime\prime},\delta_{3}^{\prime}\right]  \left(  \left(
N,a\right)  ,\left(  \mu,a\right)  ,\left(  \varkappa,a\right)  \right)
=\delta_{3}^{\prime\prime\prime}\left(  \left(  N,a\right)  ,\left(
\mu,a\right)  ,\left(  \varkappa,a\right)  \right)  ,\text{ \ \ \ \ \ }%
\theta_{3}^{\prime\prime\prime}=\dot{\theta}_{3}^{\prime\prime}\theta
_{3}^{\prime}-\dot{\theta}_{3}^{\prime}\theta_{3}^{\prime\prime}%
~,\label{eqnS50}%
\end{equation}
which is simpler than (\ref{eqnS38}), but not as simple as the `incorrect'
(\ref{eqnS35}).

The EPS approach, designed to fix the `failure' of Dirac's method, was applied
to this simple model in section 4 of \cite{ShestakovaCQG} to illustrate its
advantages. \ Long calculations were performed and `for the original variable
$N$ one gets the transformation [(S33)]' -- the `second correct'
transformation. This outcome demonstrates that the author's approach is not a
reliable algorithm. \ It focuses on an \textit{a priori} `canonization' of one
symmetry ($\delta_{2}$); yet the author's method is insensitive to this choice
of symmetry; instead, it successfully produces $\delta_{3}$ without contradiction.

A second advantage of the proposed method, emphasized by the author of
\cite{ShestakovaCQG}, is that it supports the canonical structure of Poisson
Brackets (PBs) in the \textit{extended phase space}. For the three
parametrisations considered, $\left(  N,\pi_{N}\right)  $, $\left(  \mu
,\pi_{\mu}\right)  $,\ $\left(  \varkappa,\pi_{\varkappa}\right)  $ (the
second pair, $\left(  a,p\right)  $, is decoupled), all PBs are canonical
\textit{without EPS} because of the following relations:%

\[
N=\sqrt{\mu},~~\pi_{N}=2\sqrt{\mu}\pi_{\mu}\text{ and \ }N=e^{-\varkappa
},~~\pi_{N}=-e^{\varkappa}\pi_{\varkappa}\text{\ }.
\]

The canonicity of the PBs is a necessary, but not sufficient condition for
there to be an equivalence of constrained Hamiltonians (see \cite{Myths,
FKK}). The EPS approach is based on the choice of a `correct' symmetry. And if
it is known, then there is no need to use Dirac's method to confirm it. But if
a symmetry is unknown, then Dirac's method (because it is
parametrisation-dependent) allows one to find different symmetries of a
Lagrangian and, at the same time, find the simplest, `canonical', symmetry
(not necessarily the `canonized' one) and the unique `canonical
parametrisation' for which the Hamiltonian gives this symmetry. For example,
if one were to apply Dirac's approach to $L_{2}$ or $L_{3}$ to find the
simplest parametrisation, one would uniquely obtain $L_{1}$. Of course, for
this simple model, this outcome is obvious from the inspection of the fields
in the Lagrangian; the presence of fields in combinations $\sqrt{\mu}$ or
$e^{-\varkappa}$ naturally suggests calling them $N$. Such a redefinition
gives the natural parametrisation for this model; and the Hamiltonian will
lead to the simplest symmetry, the symmetry that was rejected in
\cite{ShestakovaCQG}.

For more complicated theories the simplest reparametrisation is not obvious,
and a search can be difficult. Consider the ADM Lagrangian without any
\textit{a priori} knowledge of gauge symmetry (`correct' or `incorrect').
Application of Dirac's method leads to transformations that do not form a
group, making it necessary to find other parametrisations. This procedure can
be formulated as an algorithm (we did not guess $L_{3}$). We have not applied
it to the ADM Lagrangian; but we conjecture it will lead to a unique symmetry
and a unique parametrisation: diffeomorphism invariance, and the metric tensor
in which the EH action was originally written. Of these variables, which is
more `preferable because of its geometrical interpretation'
\cite{ShestakovaCQG}: Einstein's metric or the ADM variables?

In the author's article the application of the EPS approach to GR is
incomplete (the equivalence of the effective and EH actions is not
demonstrated, and the Hamiltonian not given); and the only result that was
obtained is a proof of the canonical relations of PBs for the class of
parametrisations (S9) and (S73). Again, this does not require EPS variables;
and for the case of ADM variables, this result was given in \cite{Myths} to
illustrate that having canonical PBs is not sufficient for two Hamiltonians to
be equivalent.

\section{Acknowledgment}

We would like to thanks A.M. Frolov, L.A. Komorowski, D.G.C. McKeon, and A.V.
Zvelindovsky for discussions.

\end{document}